\documentclass[aps,prl,twocolumn]{revtex4-1}

\usepackage{graphicx}
\usepackage{amsmath}
\usepackage{amssymb}
\usepackage{hyperref}
\usepackage[utf8]{inputenc}
\usepackage{mathtools}
\usepackage[english]{babel}
\usepackage{lipsum}
\hypersetup{colorlinks=true, linkcolor=blue, citecolor=blue, urlcolor=blue}

\newcommand{\ket}[1]{\left| #1 \right>} 

\newcommand{\fref}[1]{\text{Fig.}~\ref{#1}}
\newcommand{\ffref}[1]{\text{Figs.}~\ref{#1}}
\newcommand{\eref}[1]{\text{Eq.}~\eqref{#1}}
\newcommand{\eeref}[1]{\text{Eqs.}~\eqref{#1}}

\newcommand{\proj}[1]{|#1\rangle\langle #1|}

\newcommand{\Tr}{\operatorname{Tr}}

\begin{document}
\title{Unraveling the quantum nature of atomic self-ordering in a ring cavity}
\author{Stefan Ostermann}
\email{stefan.ostermann@uibk.ac.at}
\affiliation{Institut f\"ur Theoretische Physik, Universit\"at Innsbruck, Technikerstra{\ss}e~21a, A-6020~Innsbruck, Austria}
\author{Wolfgang Niedenzu}
\affiliation{Institut f\"ur Theoretische Physik, Universit\"at Innsbruck, Technikerstra{\ss}e~21a, A-6020~Innsbruck, Austria}
\author{Helmut Ritsch}
\affiliation{Institut f\"ur Theoretische Physik, Universit\"at Innsbruck, Technikerstra{\ss}e~21a, A-6020~Innsbruck, Austria}

\begin{abstract}
  Atomic self-ordering to a crystalline phase in optical resonators is a consequence of the intriguing non-linear dynamics of strongly coupled atom motion and photons. Generally the resulting phase diagrams and atomic states can be largely understood on a mean-field level. However, close to the phase transition point, quantum fluctuations and atom--field entanglement play a key role and initiate the symmetry breaking.  Here we propose a modified ring cavity geometry, in which the asymmetry imposed by a tilted pump beam reveals clear signatures of quantum dynamics even in a larger regime around the phase transition point. Quantum fluctuations become visible both in the dynamic and steady-state properties. Most strikingly we can identify a regime where a mean-field approximation predicts a runaway instability, while in the full quantum model the quantum fluctuations of the light field modes stabilize uniform atomic motion. The proposed geometry thus allows to unveil the ``quantumness'' of atomic self-ordering via experimentally directly accessible quantities.
\end{abstract}

\maketitle

\emph{Introduction.}---Coupling of an individual two-level quantum emitter to a single electromagnetic field mode displays fundamental principles of light and matter interaction. Reaching the strong coupling regime where the energy exchange between atom and field dominates environmental coupling and loss opened the research direction commonly known as cavity quantum electrodynamics~\cite{brune1987realization,raimond2001manipulating,meystre1988very,kimble1998strong}. Strong coupling between single atoms and the resonator modes induces non-linear field dynamics even on the single photon level and provides a seminal tool to study and reveal intriguing quantum effects such as superpositions, entanglement and measurement back-action in light-matter interaction~\cite{maunz2004cavity,hennrich2005transition,brooks2012nonclassical,dilley2012single,mekhov2012quantum,wickenbrock2013collective,reiserer2015cavity}.

\par

Extending this to whole ensembles of cold atoms interacting with optical resonator modes opens the domain of collective effects and light-induced long-range interactions. When one includes atomic motion the light-induced forces on laser illuminated particles in an optical resonator lead to self-ordering of the particles~\cite{domokos2002collective}. The underlying phase transition from a homogeneous density to an atom crystal bound by light has been first experimentally seen with thermal atoms~\cite{black2003observation} and more recently with Bose-Einstein condensates~\cite{baumann2010dicke,kessler2014steering}.  As the specific emerging order depends on the pump and cavity geometry~\cite{habibian2011quantum,habibian2014stationary,piazza2014umklapp,vaidya2018tunable} this opened promising possibilities for analog simulation of quantum phase transitions~\cite{baumann2010dicke}, spontaneous symmetry breaking and artificial quantum matter~\cite{leonard2017supersolid,leonard2017monitoring, mivehvar2018driven}. Recent theoretical~\cite{mivehvar2017disorder,mivehvar2019cavity, ostermann2019cavity, colella2019antiferromagnetic, halati2019cavity} as well as experminental~\cite{kroeze2018spinor,landini2018formation, kroeze2019dynamical, davis2019photon} advances push these possibilities towards spinor quantum matter based on multi-component quantum gases. A recent theoretical work~\cite{gietka2019supersolid} even predicts that the self-ordered state of a BEC in a ring cavity renders a promising platform for future high precision metrology.

\par

\begin{figure}
  \centering
  \includegraphics[width=0.35\textwidth]{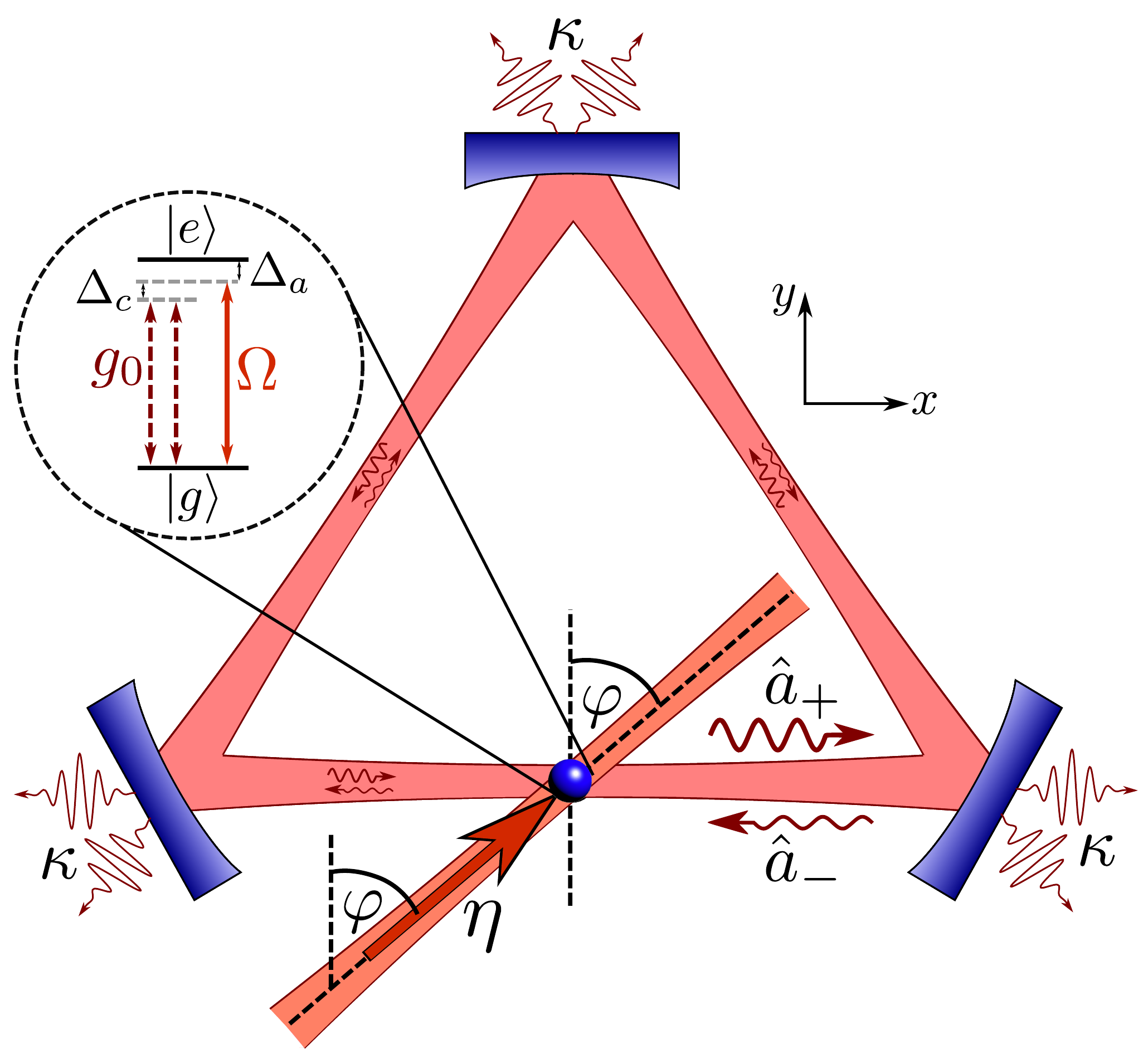} 
  \caption{Sketch of the system. The internal $\ket{g}\leftrightarrow\ket{e}$ transition of a quantum emitter moving in a ring cavity is driven at a certain angle $\varphi$ by an off-resonant plane wave laser field with pump strength $\eta$ and couples to two degenerate counter propagating cavity modes $\hat{a}_\pm$ .}
  \label{fig:setup}
\end{figure}

\par

\begin{figure}
\centering
\includegraphics[width=0.45\textwidth]{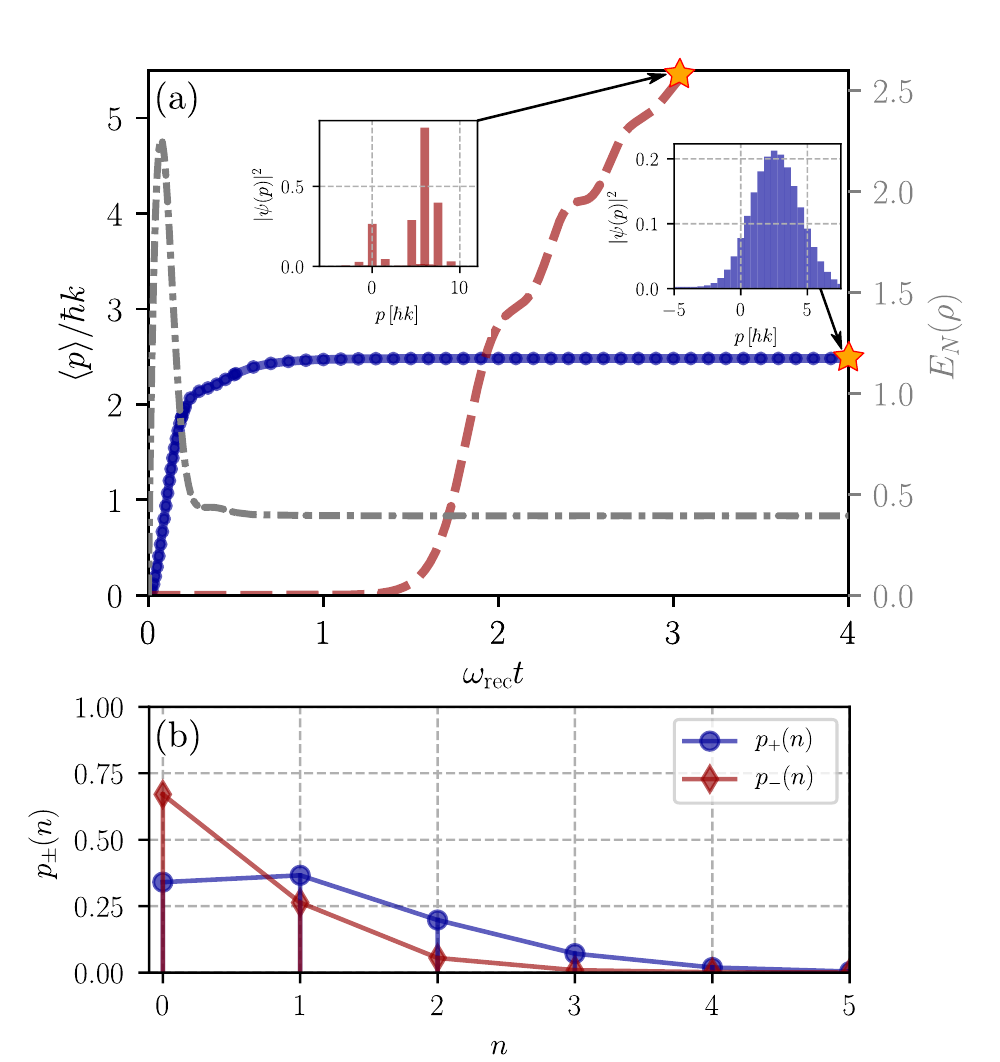} 
\caption{(a) Time evolution of the momentum expectation value $\langle\hat{p}\rangle$ (left axis) for the quantum case (solid blue line) and the mean-field model (dashed red line). The insets show the final momentum distributions for the two cases. The dash-dotted grey line (right axis) shows the time dependence of the logarithmic negativity as measure for atom--field entanglement. (b) Photon-number distribution at $\omega_\mathrm{rec}t=4$ for the quantum case (blue and red). Parameters: $(\eta,U_0,\Delta_c,\kappa)=(12,-1,-10,10)\omega_\mathrm{rec}$.}
  \label{fig:p_dyn}
\end{figure}

\begin{figure}
\centering
\includegraphics[width=0.38\textwidth]{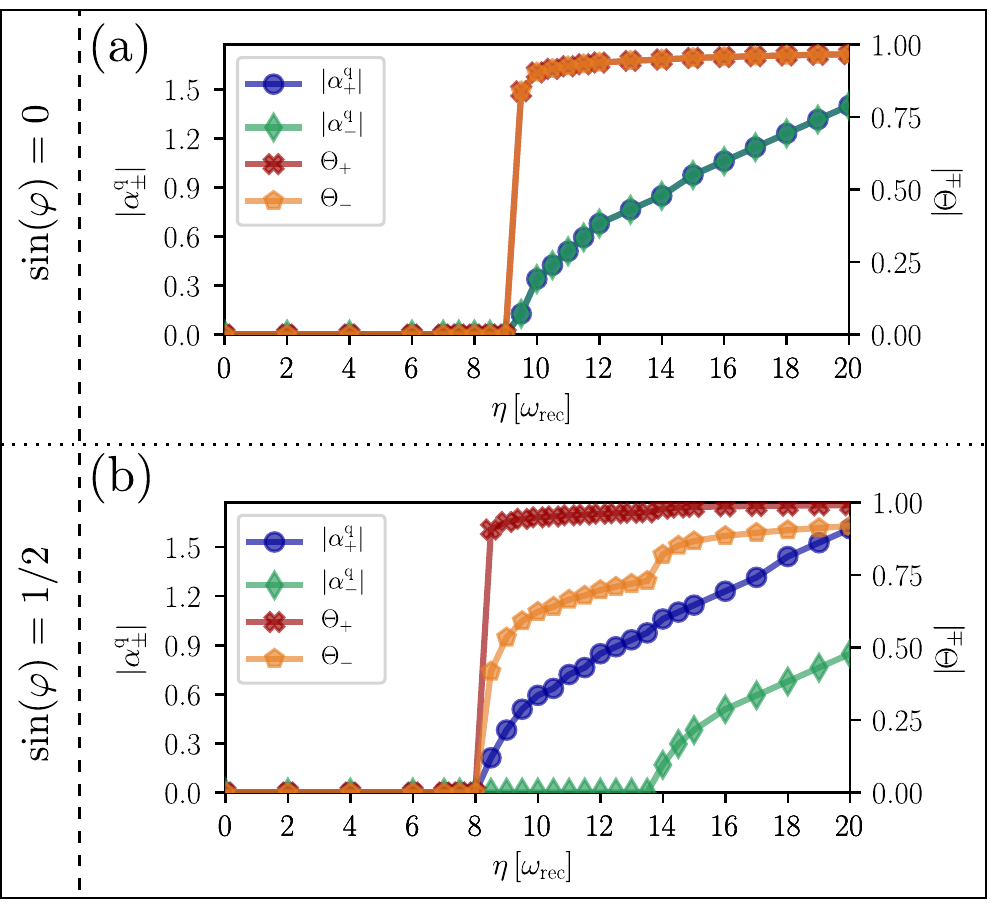}
\caption{Modulus of steady state field amplitudes and atomic order parameters as function of the pump strength $\eta$ for the full master equation quantum model  ~\eqref{eqn:master} for (a)~$\sin(\varphi)=0$ and (b)~$\sin(\varphi)=1/2$. The solid lines are a guide to the eye, the symbols mark points obtained from numerical simulations. For orthogonal pump $\sin(\varphi)=0$ the phase transition is qualitatively similar to the mean-field model discussed in~\cite{mivehvar2018driven}. For tilted pump $\sin(\varphi)=1/2$, however, the steady-state exhibits a significantly different behavior induced via photon correlations as explained in the main text. Parameters: $(U_0,\Delta_c,\kappa)=(-1,-10,10)\omega_\mathrm{rec}$.}
\label{fig:qu_vs_mf}
\end{figure}

In this letter we focus on the microscopic physics and dynamics of the self-ordering phase transition, which is closely tied to spontaneous symmetry breaking. The fundamental question addressed in the following is: How \emph{quantum} are the mechanisms behind the symmetry breaking at the onset of self-organization? Since experimental limitations and technical fluctuations currently are at least on the same order of magnitude as quantum fluctuations, it is hard to find a clear answer to this question in present experimental setups. Therefore, we propose a variation of a self-ordering setup based on a ring cavity with transversal pump~\cite{horak2001dissipative,nagy2006self,niedenzu2010microscopic,niedenzu2012quantum, sandner2015selfordered,ostermann2015atomic,colella2019antiferromagnetic,ostermann2019cavity} impinging at a \emph{non-zero} angle as shown in~\fref{fig:setup}. The fundamental modes of a ring cavity are counterpropagating running waves. Hence, the system exhibits a continuous translation symmetry, which is a crucial property for the results presented below. We state that the proposed setup allows the observation of quantum noise driven dynamics via directly accessible macroscopic observables. Our predictions are based on the comparison of a full quantum model with the mean-field dynamics. While basic properties of self-ordering of atoms in an optical resonator can be understood on a mean-field level~\cite{ritsch2013cold}, higher order density correlations or atom--field entanglement obviously cannot be properly accounted for in this model. However, these effects play a key role around threshold. Note that the fast generation of atom--field entanglement was proposed as one central mechanism driving the self-ordering phase transition in a Fabry-Pérot cavity~\cite{maschler2007entanglement,vukics2007microscopic}. We show that if the pump beam impinges onto the atoms at an angle $\varphi$ (see~\fref{fig:setup}), the mirror symmetry between the two counter-propagating cavity modes is broken, but the system still is translation invariant. This results in significant differences in the dynamics as well as the steady state properties between the mean-field and the full quantum model (see~\ffref{fig:p_dyn} and~\ref{fig:qu_vs_mf}). These differences can serve as a measure of the ``quantumness'' of a given setup via easily accessible quantities.

\emph{Quantum versus mean-field modeling.}---Let us consider a two-level atom moving along the axis of a ring resonator driven by an off-resonant plane wave laser of frequency $\omega_l$ and Rabi frequency $\Omega$ (see~\fref{fig:setup}). While the particle can freely move along the cavity axis ($x$-direction), it is strongly confined in the transverse directions. The laser,  detuned by $\Delta_a:=\omega_l-\omega_a$ from the atomic transition frequency $\omega_a$,  impinges at an angle $\varphi\in\left[-\pi,\pi\right]$. The atomic dipole couples with amplitude $g_0$ to two counterpropagating degenerate cavity modes $\hat{a}_\pm$ of the ring resonator with frequency $\omega_c$. 

In the far detuned case $|\Delta_a|\gg\Omega,g_0$ the atomic excited state can be adiabatically eliminated~\cite{ritsch2013cold} and the effective Hamiltonian in a frame rotating at the laser frequency $\omega_l$ is the sum of the atomic and the cavity Hamiltonian, $\hat{H}= \hat{H}_\mathrm{at}+\hat{H}_\mathrm{cav}$, with 
\begin{subequations}\label{eq_H}
  \begin{multline}
    \hat{H}_{\rm at}=\frac{\hat{p}^2}{2m}\\
    +\hbar U_0 \left(\hat{a}_+^\dagger \hat{a}_+ +\hat{a}_-^\dagger \hat{a}_-+\hat{a}_+^\dagger \hat{a}_- e^{-2ik\hat{x}}+\hat{a}_-^\dagger \hat{a}_+ e^{2ik\hat{x}}\right) \\
    +\hbar \eta \left(\hat{a}_+e^{ik\hat{x}(1-\sin{\varphi})}+\hat{a}_- e^{-ik\hat{x}(1+\sin{\varphi})}+ \text{H.c.}\right),
    \label{eqn:Hat}
\end{multline}
and
\begin{equation}
  \hat{H}_\mathrm{cav} =-\hbar\Delta_c\left(\hat{a}_+^\dagger \hat{a}_++\hat{a}_-^\dagger \hat{a}_-\right).
  \label{eqn:Hcav}
\end{equation}
\end{subequations}
Here we introduced the atomic momentum operator $\hat{p}=- i\hbar\partial_x$, the cavity potential depth per photon $\hbar U_0:=\hbar g_0^2/\Delta_a$, the effective pump amplitude $\hbar \eta:=\hbar\Omega g_0/\Delta_a$ from scattering of pump photons into the cavity modes with wave number $k=c\omega_c =2\pi/\lambda$, where $\lambda$ is the cavity resonance wavelength, and the cavity detuning $\Delta_c:=\omega_l-\omega_c$. The model~\eqref{eq_H} assumes a strong transversal confinement of the atom. Based on a closely-related recent experiment~\cite{wolf2018observation}, we are confident that this condition can be relaxed as long as the atom does not experience significant transverse intensity gradients on experimental time scales.

For a perpendicular pump direction, i.e., $\varphi=0$, the Hamiltonian~\eqref{eq_H} reduces to the model studied in, e.g., Refs.~\cite{nagy2006self, mivehvar2018driven}, which above a certain critical pump strength exhibits a phase transition to a supersolid state by breaking a continuous translation symmetry. Pumping at an angle $\varphi$ in Eqs.~\eqref{eq_H} preserves the continuous symmetry, i.e., $\hat{H}$ is invariant under spatial translations $\hat{x}\mapsto \mathcal{T}_{\Delta \hat{x}}\hat{x}=\hat{x}+\Delta \hat{x}$ since those are compensated by phase shifts $\hat{a}_\pm \mapsto \mathcal{U}_{\Delta x}\hat{a}_\pm=\hat{a}_\pm e^{\mp ik\Delta \hat{x}}$ of the cavity modes.

Using standard quantum optics modeling the dynamics of the composite atom-cavity system is governed by the master equation~\cite{wallsbook}
\begin{equation}
\dot{\rho}=-\frac{i}{\hbar}[\hat{H},\rho]+\kappa\sum_{j=\pm} \left(2\hat{a}_j\rho \hat{a}_j^\dagger-\hat{a}_j^\dagger\hat{a}_j \rho-\rho \hat{a}_j^\dagger\hat{a}_j\right),
\label{eqn:master}
\end{equation}
where photon loss out of the cavity at rate $2\kappa$ is included but atomic spontaneous emission is neglected~\cite{maschler2005cold}.

\par

As mentioned above, important aspects of self-ordering can already be analyzed on a mean-field level~\cite{ritsch2013cold}, where the photon field operators in Eqs.~\eqref{eq_H} are replaced by their expectation values $\hat{a}_\pm\rightarrow \langle \hat{a}_\pm\rangle\eqqcolon\alpha_\pm^\mathrm{mf}=|\alpha_\pm^\mathrm{mf}|\exp(i\phi_\pm^\mathrm{mf})$, resulting in an effective atomic Hamiltonian $\hat{H}_\mathrm{mf}\coloneqq\hat{H}_\mathrm{at}\vert_{\hat{a}_\pm\mapsto\alpha_\pm^\mathrm{mf}}=\hat{p}^2/(2m)+V_\mathrm{mf}(x)$ with the classical optical potential
\begin{align}
  V_\mathrm{mf}(\hat{x})&=2\hbar U_0 |\alpha_+^\mathrm{mf}||\alpha_-^\mathrm{mf}|\cos(2k\hat{x}+\Delta\phi^\mathrm{mf})\notag\\
  +2\hbar\eta\Big[&|\alpha_+^\mathrm{mf}|\cos\Big(k\hat{x}(1-\sin(\varphi))+\phi_+^\mathrm{mf}\Big)\notag\\
  +&|\alpha_-^\mathrm{mf}|\cos\Big(k\hat{x}(1+\sin(\varphi))-\phi_-^\mathrm{mf}\Big)\Big].
     \label{eqn:Vmf}
\end{align}
Here $\Delta \phi^\mathrm{mf}\coloneqq\phi_+^\mathrm{mf}-\phi_-^\mathrm{mf}$ denotes the relative phase between the two counterpropagating modes. Note that owing to the continuous symmetry of $\hat{H}$ the phases $\phi_\pm^\mathrm{mf}$ in the ordered phase can take arbitrary values and the dynamics is then described by the three coupled mean-field equations for the atomic wave-function $\psi(\hat{x},t)$ and the mean field amplitudes $\alpha_\pm^\mathrm{mf}(t)$,
\begin{subequations}\label{eqn:mf_dyn}
  \begin{align}
    i\hbar\partial_t \psi&=\hat{H}_\mathrm{mf}\psi \\
    i\partial_t \alpha_\pm^\mathrm{mf}&=\left(-\Delta_c+U_0-i\kappa \right)\alpha_\pm^\mathrm{mf}+U_0 \mathcal{B}_\pm^*\alpha_\mp^\mathrm{mf}+\eta \Theta_\pm^*.
  \end{align}
\end{subequations}
Here we introduced the bunching parameter $\mathcal{B}_\pm:=\langle e^{\pm 2ik\hat{x}}\rangle_\psi$ and the order parameters $\Theta_\pm:=\langle e^{\pm ik\hat{x}(1\mp\sin(\varphi))}\rangle_\psi$.

\par

To unravel the role of quantum effects in this system, we calculate the dynamics of the master equation~\eqref{eqn:master} and compare it to the mean-field dynamics obtained from Eqs.~\eqref{eqn:mf_dyn}. We numerically solve Eqs.~\eqref{eqn:master} and~\eqref{eqn:mf_dyn} using the QuantumOptics.jl framework~\cite{kraemer2018quantumoptics}. For $\varphi=0$ the Hamiltonian~\eqref{eq_H} is $\lambda$-periodic with unit cell $x\in[0,\lambda]$. For angles $\varphi\neq 0$ the periodicity of the Hamiltonian depends on $\varphi$. While certain choices of $\varphi$ may generate infinitely large unit cells, restricting the angle to values where $\sin(\varphi)$ is a rational number $n/m$ with $(n,m)\in\mathbb{Z}$ results in unit cells whose length is given by the least common multiple (LCM) of $n$ and $m$. Therefore, w.l.o.g.\ we restrict our discussion to $\varphi=\pi/6$, i.e., $\sin(\varphi)=1/2$, resulting in a unit cell $x\in[0,2\lambda]$.


\par

\par

\emph{Dynamics.}--- The dynamics close to the phase transition point allows to extract experimental signatures about the ``quantumness'' of the system.  Allowing the transverse pump field to impinge at an angle $\varphi$ generates significant differences between the two models~\eqref{eqn:master} and~\eqref{eqn:mf_dyn} for a large parameter range and reveals the quantum nature of the self-ordering phase transition close to the critical point. In~\fref{fig:p_dyn}(a) we compare the momentum expectation values obtained from the respective dynamics of~\eeref{eqn:master} (solid blue curve) and~\eqref{eqn:mf_dyn} (dashed red curve) for $\sin(\varphi)=1/2$. In either case, the tilted pump beam causes a non-zero momentum. However, while in the quantum case the atom attains a momentum constant in time, the mean-field treatment reveals a constant acceleration, i.e., an increasing momentum. These differences are also expressed in the respective momentum distributions [\fref{fig:p_dyn}(a)]: In the quantum case a broad distribution around the mean momentum is found, whereas in the mean-field case only a single momentum component is dominant.

The dynamic instability obtained in the mean-field case is well-known as the CARL (collective atomic recoil lasing) instability, which was rigorously studied theoretically and experimentally~\cite{bonifacio1994exponential,bonifacio1994collective,kruse2003observation,slama2007cavity}. Its origin is the classical nature of cavity fields in the mean-field treatment: There is no photon number distribution and therefore one mode always contains more photons than the other. This imbalance causes the runaway CARL instability.

\par

By contrast, the quantum nature of the cavity fields suppresses the CARL runaway effect in the quantum case: There is a significant propability that no photon is scattered in the mode $\hat{a}_+$ but a photon is scattered in the mode $\hat{a}_-$ [\fref{fig:p_dyn}(b)]. Hence, the photon statistics is more balanced than in the mean-field case. This is also reflected in the broad atomic momentum distribution [\fref{fig:p_dyn}(a)] which even contains a significant amount of negative momentum components. Note that the monotonically-decreasing photon number distribution of the mode $\hat{a}_-$ shows that the state is a so-called \emph{passive state}~\cite{pusz1978passive,lenard1978thermodynamical} whose energy cannot be reduced by cyclic unitary transformations. Passive states other than the vacuum state only exist if that state has a finite entropy. Here part of this entropy is generated by partially tracing over a correlated atom--fields state. The distribution for the mode $\hat{a}_+$ is non-monotonic and therefore the state is non-passive. By contrast, in the mean-field treatment the field states are implicitly assumed to be non-passive.

\par

The role of atom--field entanglement for the dynamics of the quantum case can be analyzed by calculating the logarithmic negativity~\cite{vidal2002computable}. For the bipartite system consisting of the subsystems $A$ (atoms) and $B$ (modes $\hat{a}_\pm$) it is defined as $E_N(\rho)=\log_2(\|\rho^{T_A}\|)$, where $\rho^{T_A}$ denotes the partial transpose with respect to the subsystem $A$ and $\|\rho\|\coloneqq \Tr(\sqrt{\rho^\dagger\rho})$ is the trace norm. The entanglement increases to very high values as long as the momentum expectation value increases [grey dashed-dotted line in~\fref{fig:p_dyn}(a)] and it saturates as soon as the steady state is reached. Hence, atom--field entanglement plays a major role during the build up phase of the moving lattice.

\par

In summary, the dynamics of an atom in a ring cavity with non-perpendicular transversal pump allows to directly observe the effect of quantum statistics. If the mean-field treatment describes the system well one would observe a runaway CARL instability. However, as soon as the quantum nature of the constituents start to play a role one would observe an atom moving with a constant center of mass velocity. This makes this setup a seminal tool to study quantum effects in the self-ordering phase transition.

\par

\emph{Steady-state properties.---} Besides the difference in the dynamics, the effect of quantum nature also appears in the steady state of the master equation~\eqref{eqn:master}. As a result of the continuous symmetry of the Hamiltonian $\hat{H}$ the full quantum steady state density matrix  $\rho_\mathrm{ss}$ of~\eref{eqn:master} exhibits the same symmetry. It thus contains all states which can be transformed into each other via spatial translations $\mathcal{T}_{\Delta x}$ and corresponding phase shifts $\mathcal{U}_{\Delta x}$. Therefore, in the steady state the average order parameters $\Theta_\pm$ and field amplitudes $\alpha_{\pm}$ vanish~\cite{sandner2015selfordered,gietka2019supersolid} as the dynamics contains no process that spontaneously breaks the system's continuous symmetry.

However, selecting a particular field phase unveils the constituents of the corresponding atom--field state. To this end we introduce the an effective potential $V_\mathrm{quant}\coloneqq V_\mathrm{mf}\vert_{(\alpha_\pm^\mathrm{mf},\phi_\pm^\mathrm{mf})\mapsto(\alpha_\pm^\mathrm{q},\phi_\pm^\mathrm{q})}$ with the absolute values $|\alpha_\pm^\mathrm{q}|$ of the cavity field amplitudes obtained from the maxima of the Wigner functions~\cite{wallsbook} $W_\pm$ of the field modes states $\rho_\pm\coloneqq\Tr_{\mathrm{at},\hat{a}_\mp}(\rho_\mathrm{ss})$ (see also Ref.~\cite{sandner2015selfordered}). By choosing specific phases $\phi_\pm^\mathrm{q}$ the symmetry is broken explicitly. Note that the potential $V_\mathrm{quant}$ obeys the same periodicity as the initial Hamiltonian~\eqref{eq_H}. The ground state of the newly found Hamiltonian $\hat{\tilde{H}}=\hat{p}^2/2m+V_\mathrm{quant}(\hat{x})$ is a symmetry broken state with non-vanishing order parameters. The comparison of the resultant states for $\sin(\varphi)=0$ and $\sin(\varphi)=1/2$ unravels the role of quantum effects in the self-ordering process close to the transition point.

\par

\par

\begin{figure}
\centering
\includegraphics[width=0.45\textwidth]{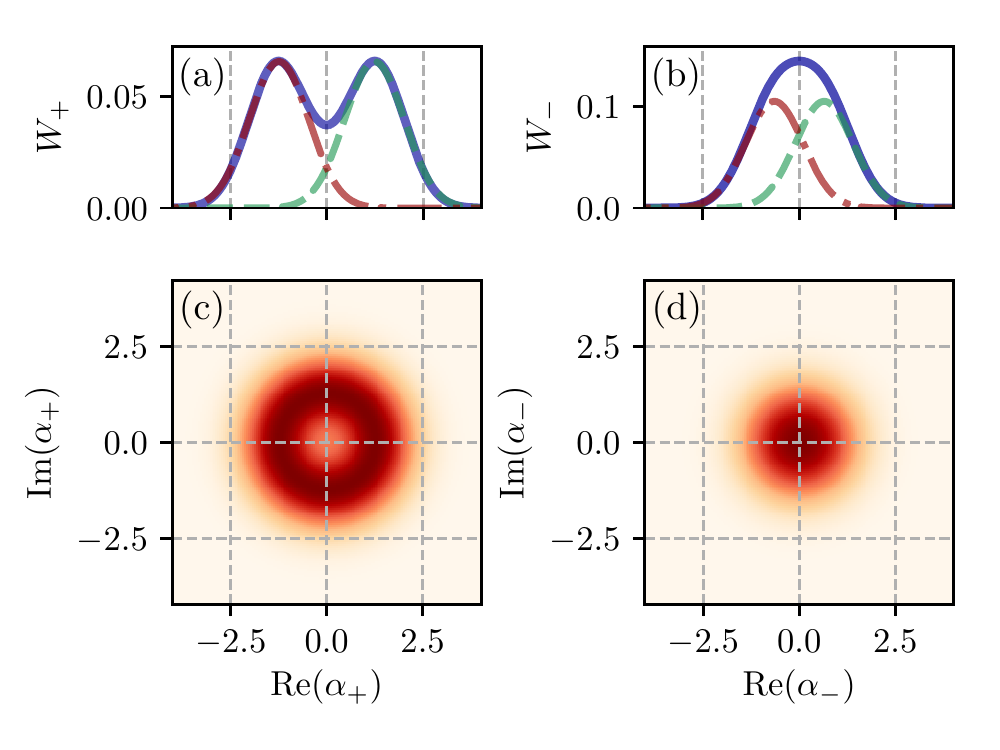}
\caption{Wigner functions $W_\pm$ of the field modes $\hat{a}_+$ and $\hat{a}_-$ for $\sin(\varphi)=1/2$ and $\eta=12\omega_\mathrm{rec}$ for parameters as in~\fref{fig:qu_vs_mf}. (a) and (b): radial distribution obtained by cutting $W_\pm$ [see (c) and (d)] along $\mathrm{Im}(\alpha_\pm)=0$. Whereas the Wigner function $W_+$ of the mode $\hat{a}_+$ has a clear central minimum, $W_-$ of the mode $\hat{a}_-$ has a center maximum.}
\label{fig:Wfunc}
\end{figure}

\fref{fig:qu_vs_mf} shows the comparison of the cavity field amplitudes and order parameters $\Theta_{\pm}$ for $\sin(\varphi)=0$ [\fref{fig:qu_vs_mf}(a)] and $\sin(\varphi)=1/2$ [\fref{fig:qu_vs_mf}(b)] as a function of the pump strength $\eta$. The two cases exhibit significantly different behavior. In the traditional case all order parameters and modes exhibit a clear threshold behavior at the same critical pump strength. For the case where the transversal pump beam impinges non-perpendicular to the cavity axis [\fref{fig:qu_vs_mf}(b)], however, the field $|\alpha_+^\mathrm{q}|$ and the corresponding order parameters $\Theta_\pm$ have a different threshold than the field $|\alpha_-^\mathrm{q}|$. Hence, there is a region where the $\hat{a}_+$ mode has a field $|\alpha_+^\mathrm{q}|\neq 0$ whereas the $\hat{a}_-$ mode still has no field, $|\alpha_-^\mathrm{q}|=0$. To understand the strong discrepancy between the two cases we analyze the properties of the quantum state obtained from~\eref{eqn:master} in more detail. \fref{fig:Wfunc} shows the Wigner functions $W_\pm$ \cite{wallsbook} of the modes $\hat{a}_\pm$ for pump strength $\eta=12\omega_\mathrm{rec}$ and $\sin(\varphi)=1/2$, which corresponds to the aforementioned region where the mode $\hat{a}_-$ has no field in contrast to the mode $\hat{a}_+$. While $W_+$ has the form of an annulus [\fref{fig:Wfunc}(c)], $W_-$ exhibits a single peak at the origin [\fref{fig:Wfunc}(d)]. Note that the rotational symmetry of the Wigner functions is a direct result of the continuous translational symmetry of the system (see also Ref.~\cite{gietka2019supersolid}).

\par

As mentioned above, we use the location of the maxima of $W_\pm$ to determine the magnitude of the field amplitude of the respective mode. Hence, we only attribute a non-zero field to the mode $\hat{a}_\pm$, if the radial cut of $W_\pm$ exhibits two equal maxima, which requires $W_\pm$ to have the form of an annulus [see~\ffref{fig:Wfunc}(a)--(b)]. While this fixes $|\alpha_\pm^\mathrm{q}|$, the phase $\phi_\pm^\mathrm{q}$ still remains unspecified and may be chosen arbitrarily. This property is also found in traditional laser setups~\cite{wiseman1997defining,scullybook}. As shown in~\fref{fig:Wfunc}, the reduced photon states are phase-averaged coherent states (Poissonian states) whose radial distribution of the Wigner representation is the sum of two Gaussian distributions. The Wigner function of the phase-averaged coherent state $\rho_\lambda=\exp(-\lambda^2)\sum_n \frac{\lambda^{2n}}{n!}\proj{n}$ ($\lambda>0$) exhibits multiple maxima (uniformly distributed on a ring) iff $\lambda>1/2$, which is the quantum noise limit.

\par

\par

For the chosen pump strength $\eta=12\omega_\mathrm{rec}$ only the mode $\hat{a}_+$ exhibits a non-zero field amplitude $|\alpha_+^\mathrm{q}|>0$,  whereas for the $\hat{a}_-$ mode $|\alpha_-^\mathrm{q}|=0$. This is in stark contrast to the ``classical'' mean-field treatment which ignores the quantum noise and explicitly breaks the symmetry by attributing a non-zero mean-field $|\alpha_-^\mathrm{mf}|>0$ [the position of the maximum of the green dashed and red dash-dotted curve in~\fref{fig:Wfunc}(b)]. By contrast, in the quantum treatment quantum effects and correlations prevent the emergence of a symmetry-breaking field in that mode. This implies that for $\sin(\varphi)=0$ the quantum nature of the fields only plays a role in a very narrow region around threshold. Including an angle $\sin(\varphi)\neq 0$, however, results in a wider region where the the system's quantum nature in revealed.
 
\emph{Conclusions.}---The comparison of a full quantum description with a mean-field model for the self-ordering of atoms in a ring resonator with a tilted transverse pump beam reveals the strong role of quantum fluctuations in the corresponding phase transition. Breaking mirror symmetry by introducing an angle for the pump light creates significant differences in both, the dynamics as well as the steady-state properties. Striking dynamical differences arise from the quantum fluctuations and quantum correlations (entanglement) of cavity fields and atoms neglected in the mean-field treatment. Hence, the system provides a clear tool to measure the ``quantumness'' of a given setup via easily accessible quantities. This is of particular experimental relevance since the study of quantum effects in the self-ordering phase transition is in general difficult as they are often hidden due to technical noise. Our setup is realizable with only minor modifications to current experiments~\cite{schneeweiss2017fiber,cox2018increased,naik2018bose,wolf2018observation,johnson2019observation} and should also be relevant in quantum thermodynamics, where the fundamental difference between passive and non-passive states plays a key role~\cite{allahverdyan2004maximal,vinjanampathy2016quantum,brown2016passivity,niedenzu2018quantum}.

\emph{Acknowledgments}---We would like to thank Karol Gietka and Farokh Mivehvar for fruitful discussions. W.\,N.\ acknowledges support from an ESQ fellowship of the Austrian Academy of Sciences (\"OAW). S.\,O.\ acknowledges support from the FWF project I 3964-N27.

\end{document}